\documentclass[pra,aps,twocolumn,amsmath,amssymb,floatfix,pra,reprint,footinbib,superscriptaddress,longbibliography]{revtex4-2}
\usepackage{graphics}      
\usepackage{graphicx}      
\usepackage{longtable}     
\usepackage{url}           
\usepackage{bm}    
\usepackage{braket}        
\usepackage{xcolor}
\usepackage{float}  
\usepackage{natbib}
\usepackage{graphicx}
\usepackage{comment}
\usepackage{siunitx}
\usepackage{appendix}
\usepackage{amsmath,amssymb,epsfig,color}

\usepackage[most]{tcolorbox}

\newcommand{\nn}{\nonumber}
\newcommand{\nl}{\nonumber \\}
\newcommand{\be}{\begin{equation}}
\newcommand{\ee}{\end{equation}}
\newcommand{\bea}{\begin{eqnarray}}
\newcommand{\eea}{\end{eqnarray}}

\begin{document}

\title {Synthesizing Arbitrary Non-Hermitian Hamiltonian with Stochastic Floquet Engineering}

\author{Lingzhen Guo}
\affiliation{College of Science, NUDT, Changsha, 410073, China}
\affiliation{Center for Joint Quantum Studies and Department of Physics, School of Science, Tianjin University, Tianjin 300072, China }

\author{Hui Jing}
\thanks{jinghui@nudt.edu.cn}
\affiliation{College of Science, NUDT, Changsha, 410073, China}
\affiliation{Department of Physics, Hunan Normal University, Changsha, 410081, China}

\begin{abstract}
The conventional Floquet engineering scheme synthesizes a given target Hamiltonian with a deterministic temporal periodic driving field. In this work, we introduce the stochastic Floquet engineering scheme that can synthesize an arbitrary non-Hermitian target Hamiltonian using a time-periodic driving field with noisy amplitude. Our method is rooted in the Hermitian dynamics taking noise as a valuable quantum resource with no need for loss or gain in prior. We apply our method to engineer a cavity Hamiltonian with dissipative coupling between Fock states, and to prepare a given quantum state from a generally arbitrary quantum state. The stochastic Floqut engineering also provides a way to generate non-unitary quantum gates, which take advantage in certain tasks compared to unitary quantum computing, without the need for ancillae or state-dependent updating. 
\\
\end{abstract}

\date{\today}

\maketitle


\section{Introduction}\label{}


Since the seminal discovery of parity-time ($\mathcal{PT}$) symmetry~\cite{bender1998prl,bender2024rmp}, which allows certain non-Hermitian (NH) operators to have real eigenvalues, the study of NH physics has grown rapidly in the last decades~\cite{ganainy2018np,ashida2020aip}. A plethora of exotic NH phenomena have been discovered, such as $\mathcal{PT}$-phase transitions~\cite{lv2015prl,liu2016prl,yang2019np,zhang2020nano,weidemann2022nature}, NH skin effects~\cite{zhang2022apx}, NH topology~\cite{okuma2019prl,longhi2019prl,zhang2022nanoph,okuma2023arcmp}, and the physics of exceptional points (EPs)~\cite{lv2017prapp,jing2017sr,lvhao2018prapp,huang2022lpr,zhang2022nc,bergholtz2021rmp,xing2025arxiv}. Various NH Hamiltonians have been implemented on different experimental platforms, e.g., solid-state systems~\cite{yang2019np}, photonic structures~\cite{regensburger2012nature}, ion traps~\cite{wu2025arxiv}, and optomechanical systems~\cite{doppler2016nature,xu2016nature}. However, most works focus on the special type of NH Hamiltonians, such as $\mathcal{PT}$-symmetric Hamiltonians and dissipative Hamiltonians. It is of great interest to have a general scheme that can engineer arbitrary NH Hamiltonians in physical systems. 

The time evolution of a quantum system subjected to an NH Hamiltonian is described by a non-unitary transformation. A growing interest was recently drawn to implementing direct non-unitary operations~\cite{terashima2005ijoqi} for quantum technologies, such as quantum steering~\cite{cavalcanti2017rpp,uola2020rmp,roy2020prr}, measurement-induced entanglement transitions~\cite{li2018prb,chan2019prb,skinner2019prx,fux2024prr}, block encoding~\cite{harrow2009prl,gilyen2019acm,biswa2024prr,leadbeater2024qst}, and the imaginary time evolution~\cite{motta2020np,sun2021prxq,grundner2024prb}.
While conventional quantum computers are built based on unitary quantum gate operations, direct non-unitary operations take advantage in certain tasks, e.g., computing the ground state of a Hamiltonian, thermal average of operators, and the dynamics of open quantum systems~\cite {watad2024qst,lin2021prxq}. The non-unitary quantum circuit requires fewer qubits to perform some kinds of quantum computation~\cite{terashima2005ijoqi,zhang2025prl}. It is also widely believed that the non-unitary quantum computer can solve some $NP$-complete problems in polynomial time, which cannot be done with a unitary quantum computer~\cite{abrams1998prl}.

\begin{figure}
  \centering
 \centerline{\includegraphics[width=1.0\linewidth]{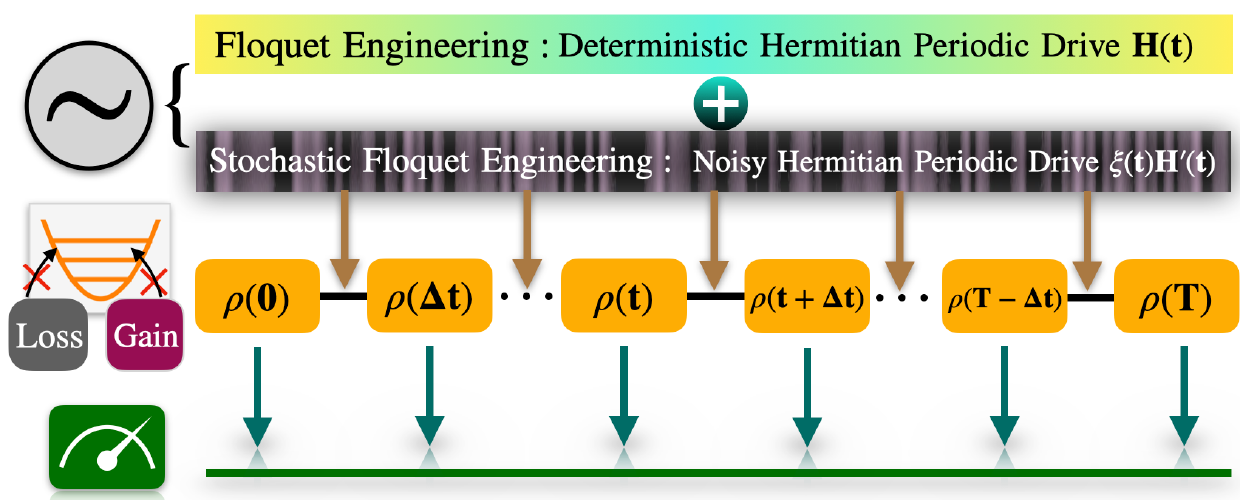}}
  \caption{{\bf Sketch of Stochastic Floquet Engineering (SFE).}
 The conventional Floquet engineering uses the deterministic Hermitian time-periodic drive $H(t)=H(t+T)$, while the SFE  adds another stochastic Hermitian time-periodic drive  $H'(t)=H'(t+T)$ with noisy amplitude $\xi(t)$. By monitoring quantum jumps, any target non-unitary time evolution of the density operator  $\rho(T)=e^{-iH_TT}\rho(0)e^{iH^\dagger_TT}$ generated by an arbitrary non-Hermitian(NH) Hamiltonian $H_T\neq H^\dagger_T$  can be realized with no need for loss and gain in prior.
   }\label{Fig-sketch}
\end{figure}

In general, it is impossible to implement a non-unitary quantum gate operation deterministically due to the unitary time evolution of quantum systems at the physical level~\cite{terashima2005ijoqi,leadbeater2024qst}. 
One possible solution to achieve non-unitary operations is to introduce ancillae together with postselection like the block encoding method, which relies on a larger unitary quantum circuit and selected ancillary measurements~\cite{harrow2009prl}.
It is also possible to perform non-unitary operations without ancillae but conditioned on the knowledge of quantum state~\cite{terashima2005ijoqi}, e.g, representing the normalized state of a non-unitary step by a unitary transformation updated via quantum measurements and classical feedback~\cite{motta2020np,watad2024qst}. 
In both schemes, every single non-unitary operation needs quantum measurements for either postselection or state-dependent updating. 

In this work, we introduce a general framework of \textit{stochastic Floquet engineering} (SFE), together with postselection suppressing quantum jumps, to synthesize an arbitrary target NH Hamiltonian with no need for ancillae and state-dependent updating. Also different from the conventional schemes generating NH Hamiltonians using loss or gain in prior, our SFE method is based on the Hermitian dynamics via time-periodic driving with noisy amplitude. Our scheme shows that noise can be a valuable quantum resource engineered for NH physics.

\section{Stochastic Floquet engineering}\label{}

To synthesize a given non-unitary evolution $\mathcal{U}=e^{-iH_TT}$ generated by the target NH Hamiltonian $H_T=H_R-iH_I$ over a reference time period $T$, we engineer a stochastic time-dependent Hamiltonian operator for a quantum system as follows
\bea
H_{s}(t)=H(t) + \sqrt{\eta}\xi(t)H'(t).
\eea
Here, $H(t)$ and $H'(t)$ are both Hermitian operators, and $\xi(t)$ represents the standard white noise that can be produced by a random number generator in the experiment.
Given an initial state of density matrix $\rho(t_0)$, the time evoluted state is given by $\rho(t)=U_\xi(t)\rho(t_0)U^\dagger_\xi(t)$ with 
\bea\label{eq-xit}
U_\xi(t,t_0)\equiv\mathcal{T}e^{-i\frac{1}{\lambda}\int_{t_0}^{t}\big[{H}(\tau)+\sqrt{\eta}\xi(\tau)H'(\tau)\big]d\tau}.\ 
\eea
Here, $\mathcal{T}$ represents the time-ordering operator, and $\lambda$ is the dimensionless Planck constant. 
For the conventional Floquet engineering ($\xi=0$), the single-period time evolution can be described by $U_{\xi=0}(t_0,t_0+T)\equiv e^{-i\frac{T}{\lambda}H_F(t_0)}$  with $H_F(t_0)$ the time-\textit{independent} Floquet Hamiltonian. However, such a Hermitian Floquet Hamiltonian does not exist for a finite noisy process ($\xi \neq 0$).

We require that the white noise process is much faster than the engineered Hamiltonians $H(t)$ and $H'(t)$.
Assuming that $N_{\xi}$ discretized random numbers are evenly generated during the time step $\Delta t$, the noisy amplitude of Hamiltonian is proportional to $\Delta W=\int_{t_0}^{t_0+\Delta t}\xi(\tau)d\tau\approx \sum_{k=1}^{N_{\xi}}\Delta W_k$, where $\Delta W_k\equiv \xi(t_0+k\Delta t)\Delta t/N_{\xi}$ are Gaussian random numbers with zero mean value and variance $\Delta t/N_{\xi}$. By expanding the time evolution operator $U_\xi(t_0+\Delta t,t_0)$ to the second order of $\Delta t$, we have dynamics for the density matrix $\rho(t)$ (see more detalied derivation in Appendix.~IA ) 
\bea\label{eq-drhot}
\frac{d\rho}{dt}
&=&-i\frac{1}{\lambda}[H(t)\rho-\rho H(t)]+\frac{\eta}{\lambda^2} \mathcal{L}[H'(t)](\rho)\nl
&=&-i \frac{1}{\lambda}\big[\mathcal{H}(t)\rho-\rho \mathcal{H}^\dagger(t)\big]+ \frac{\eta}{\lambda^2}H'(t)\rho H'(t).
\eea
Here, we have introduced the Lindblad term $\mathcal{L}[H'](\rho)\equiv H'\rho H'-\frac{1}{2}H'^2\rho-\frac{1}{2}\rho H'^2$,
the time-dependent NH Hamiltonian $
\mathcal{H}(t)\equiv H(t)-i\frac{1}{2}\eta H'^2(t)
$, and the the associated quantum jump term $\mathcal{J}[H'(t)](\rho)\equiv H'(t)\rho H'(t)$. 


We now set both $H(t)$ and $H'(t)$ to be time-periodic operators of a Floquet system~\cite{Floquet1883,Shirley1965pr} with periodicity $T$. By setting $t_0=0$ and discretizing time steps $t_n=n\Delta t$ with $n\in \mathbb{Z}$, the density operator after one Floquet period can be calculated from Eq.~(\ref{eq-drhot}) 
\bea\label{eq-rhoTKD}
\rho(T)
&\approx&\prod_{n=1}^{T/\Delta t} C_ne^{ \Delta t\frac{\eta}{\lambda^2}\mathcal{J}[H'(t_n)]}e^{-i\Delta t\frac{1}{\lambda} \mathcal{D}[\mathcal{H}(t_n)]}\rho(0).\ \ \ \ 
\eea
Here, $\mathcal{D}[\mathcal{H}](\rho)\equiv \mathcal{H}\rho-\mathcal{H}^\dagger\rho$ is the Hamiltonian superoperator, and $C_n$ is the normalization factor to keep the unit trace of the density operator.
 %
%
In the rotating wave approximation (RWA), the stroboscopic dynamics of the density operator over the Floquet period can be obtained from Eq.~(\ref{eq-rhoTKD}) (see Appendix.~IB)
\bea\label{eq-rhoT}
\frac{\Delta\rho}{\Delta t}
&\approx&-i\frac{1}{\lambda} \big(H_F\rho-\rho H^\dagger_F\big)+\eta\frac{1}{\lambda^2} \overline{H'\rho H'},\ \ \ 
\eea
where the time interval is set to be $\Delta t=T$, and $H_F$ is the NH Floquet Hamiltonian given by
\bea\label{eq-HF12}
H_F\equiv\overline{H(t)}-i\frac{1}{2\lambda}\eta \overline{H'^2}.
\eea
Here, the overline represents the temporal average for a time periodic operator $\overline{O(t)}\equiv T^{-1}\int_{0}^{T} {O}(t)dt$. The RWA is valid when the characteristic time scale of the Floquet Hamiltonian $H_F$ is much longer than the time period $T$. 

By setting $\eta=2\lambda$ and engineering two time-periodic Hamiltonians $ \overline{H(t)}=\lambda H_R$ and $\overline{h(t)}=\lambda H_I$, we take
\bea\label{eq-HtRI}
H'(t)=\sqrt{h(t)+c(t)I}
\eea
with $I$  the identity matrix, where $c(t)>0$ is a free gauge to guarantee the positivity of operator $h(t)+c(t)I$. Then, we have the NH Floauet Hamiltonian from Eq.~(\ref{eq-HF12})
\bea\label{eq-HFc}
H_F=\lambda(H_R-iH_I)-i\overline{c}=\lambda H_T-i\overline{c}
\eea
with $\overline{c}\equiv\overline{c(t)}$.
Note that the imaginary constant in the effective Hamiltonian (\ref{eq-HFc}) cannot be simply neglected because it results in a decay term $-2\lambda^{-1}\overline{c}\rho$ in Eq.~(\ref{eq-rhoT}). 
From Eq.~(\ref{eq-HF12}), the density matrix over one Floquet period is 
\bea\label{eq-rhoFT}
\rho(T)&=&e^{\frac{T}{\lambda}\overline{\mathcal{J}[H']}}\Big[e^{-i\frac{T}{\lambda}H_F}\rho(0)e^{+i\frac{T}{\lambda}H^\dagger_F}\Big]\nl
&=&e^{\frac{T}{\lambda}\big(\overline{\mathcal{J}[H']}-2\overline{c}\big)}\big[e^{-iTH_T}\rho(0)e^{+iTH^\dagger_T}\big],
\eea 
where we have introduced the time-averaged jump operactor $\overline{\mathcal{J}[H']}(\rho)\equiv\overline{H'\rho H'}$.

\section{Kraus summary and postselection}

With the SFE scheme, we have generated the target non-unitary time evolution $\rho(T)\propto e^{-iH_T}\rho(0) e^{iH^\dagger_T}$ embedded in Eq.~(\ref{eq-rhoFT}).
However, the target non-unitary transformation is corrupted by the quantum jumps, cf. Eqs.~(\ref{eq-rhoTKD}) and (\ref{eq-rhoFT}). 
We need to select the trajectories without quantum jumps that follow the target nonunitary time evolution. To this end, we write the elementary time evolution in Eq.~(\ref{eq-rhoTKD}) in the form of Kraus summary $\rho(t+\Delta t)=\sum_{m=0}^\infty K^\dagger_m(t)\rho(t)K_m(t),$ where the time-dependent Kraus operator is 
\bea
K_m(t)=\sqrt{\frac{1}{m!}\Big(\frac{\eta\Delta t}{\lambda^2}\Big)^m}H^{'m}(t)e^{-\frac{\eta\Delta t}{2\lambda^2}\mathcal{H}(t)}.\ \ \ 
\eea
with the completeness relationship $\sum_mK_mK^\dagger_m=I$. The Kraus set $\{K_m\}$ can also be viewed as a collection of general measurement operators, where the index $m$ refers to the measurement outcomes that may occur in the experiment~\cite{chuang2010book}. The probability that result $m$ occurs is given by $p_m(t)=\mathrm{Tr}[K_m\rho(t) K^\dagger_m]$, and the state of the system after the measurement collapses to
$
 K_m\rho(t) K^\dagger_m/p_m(t).
$
By continuously monitoring the null outcome result ($m=0$), we can realize the target non-unitary time evolution of the density matrix $\rho(T)=Ce^{-iH_T}\rho(0) e^{iH^\dagger_T}$. As we only care about the outcome $m=0$, we can reduce the measurement operators to two, i.e., $K_0(t)=e^{-\frac{\eta\Delta t}{2\lambda^2}\mathcal{H}(t)}$ and $\overline{K}_0(t)\equiv\sqrt{I-K_0(t) K^\dagger_0(t)}$.
To numerically simulate the monitoring process, one can generate a uniform random number $\zeta\in [0,1]$ during each time step $\Delta t$ and update the conditional density matrix $\rho_\zeta(t)$ of the system by
\bea\label{eq-rhozeta}
\rho_\zeta(t+\Delta t)=
\begin{cases}
  \frac{K_0(t)\rho_\zeta(t)K^\dagger_0(t)}{p_0(t)},&  \zeta\leq p_0(t)\\
     \frac{\overline{K}_0(t)\rho_\zeta(t)\overline{K}^\dagger_0(t)}{1-p_0(t)},              & \zeta>p_0(t).
\end{cases}
\eea
By generating $N_\zeta$ samples of $\rho_\zeta(t)$ trajectories, the unconditional density matrix can be approximated by the ensemble average $\rho(t)\approx\langle\rho_\zeta(t) \rangle$.
In the experiment, one can monitor a proper observable to postselect trajectories without quantum jumps that obey the target non-unitary time evolution. In Fig.~\ref{Fig-sketch}, we summarize and sketch the general framework of SFE.

\section{Bosonic NH Hamiltonian}

The SFE scheme introduced above is valid for general quantum systems. We now apply it to the bosonic system of a cavity Hamiltonian $H_0=\omega_0\left(\hat{p}^{2}+\hat{x}^{2}\right)/2$, which is subjected to one deterministic time-periodic potential $V_+(x,t)$ and another noisy-amplitude time-periodic potential $V_-(x,t)$
\begin{equation}\label{eq-HstV}
H_s(t)=H_0+\beta V_+ (\hat{x},t)+ \sqrt{2\beta}\xi(t)\sqrt{V_-(x,t)+c(t)} .
\end{equation}
Here, the time-dependent parameter $c(t)$ is the free gauge to guarantee $V_-(x,t)+c(t)\geq 0$ and $\beta$ is the driving amplitude.
We transform the above Hamiltonian into the rotating frame with time-evolution operator ${O}(t)\equiv e^{i{a}^\dagger{a}\omega_0 t}$, i.e., 
$
\tilde{H}_s(t)\equiv{O}(t)H_s(t){O}^\dagger(t)-i\lambda {O}(t)\dot{{O}}^\dagger(t).
$
Our task is to find the explicit forms for the potentials $V_\pm(x,t)$ such that the Floquet Hamiltonian corresponding to $\tilde{H}_s(t)$, cf. Eq.~(\ref{eq-HF12}),  equals the target NH Hamiltonian in the Fock basis 
$
H_T=\beta\sum_{n,m}c_{nm}|n\rangle\langle m|
$
that allows $c_{nm}\neq c^*_{mn}$ with 
real part and the imaginary parts 
\bea\label{eq-HRHI}
 \left \{ \begin{array}{lll}
H_R=\frac{1}{2}(H_T^{\dagger}+H_T)&=&\beta\sum\limits_{n,m}\frac{1}{2}\big(c^*_{mn}+c_{nm}\big)|n\rangle\langle m|\\
 H_I=\frac{1}{2i}(H^\dagger_T-H_T)&=&\beta\sum\limits_{n,m}\frac{1}{2i}\big(c^*_{mn}-c_{nm}\big)|n\rangle\langle m|.
\end{array} \right.
\eea
The driving potential $V_\pm(x,t)$ can be decomposed into a series of cosine-type lattice potentials as
\bea\label{eq-Vxt-cos}
V_{\pm}(x, t)
&=&\int_{-\infty}^{+\infty}A_{\pm}(k, t)\cos[kx+\phi_{\pm}(k, t)]dk.\ \ \ \ \ \ 
\eea
To synthesize $H_R$ and $H_I$ from $V_+(x,t)$ and $V_-(x,t)$ respectively, we adopt the non-commutative Fourier transformation (NcFT) technique~\cite{guo2024prl} that gives the tunable time-dependent amplitude  $A_{\pm}(k,t)=k\big|f^{\pm}_{T}(k,\omega_0 t)\big|$ and the phase 
 $\phi_{\pm}(k,t)=\text{Arg}\big[f^{\pm}_{T}(k,\omega_0 t)\big]$
%
with $f^{\pm}_{T}(k,\omega_0 t)$ the NcFT coefficient of the target Hamiltonian (see Appendix.~II)
\bea\label{eq-fT}
f^{\pm}_{T}(k,\omega_0 t)&=&\lambda\beta\sum_{n,m}\frac{1}{2}(c^*_{mn}\pm c_{nm}) f_{n,m}(k,\omega_0 t).\ \ \ 
\eea
Here, we have introduced the basic NcFT coefficient $f_{nm}(k,\omega_0 t)=\sqrt{\frac{n!}{m!}}\left(\frac{i}{k}\sqrt{\frac{2}{\lambda}}\right)^{m-n}\frac{\lambda e^{\frac{\lambda}{4}k^{2}+i(m-n)\omega_0t}}{\Gamma(1+n-m)}{}_{1}F_{1}(1+n; 1+n-m; -\frac{\lambda}{2}k^{2})$ with ${}_{1}F_{1}(a;b;z)$ the Kummer confluent hypergeometric function. 
%

 \begin{figure}
  \centering
 \centerline{\includegraphics[width=1.0\linewidth]{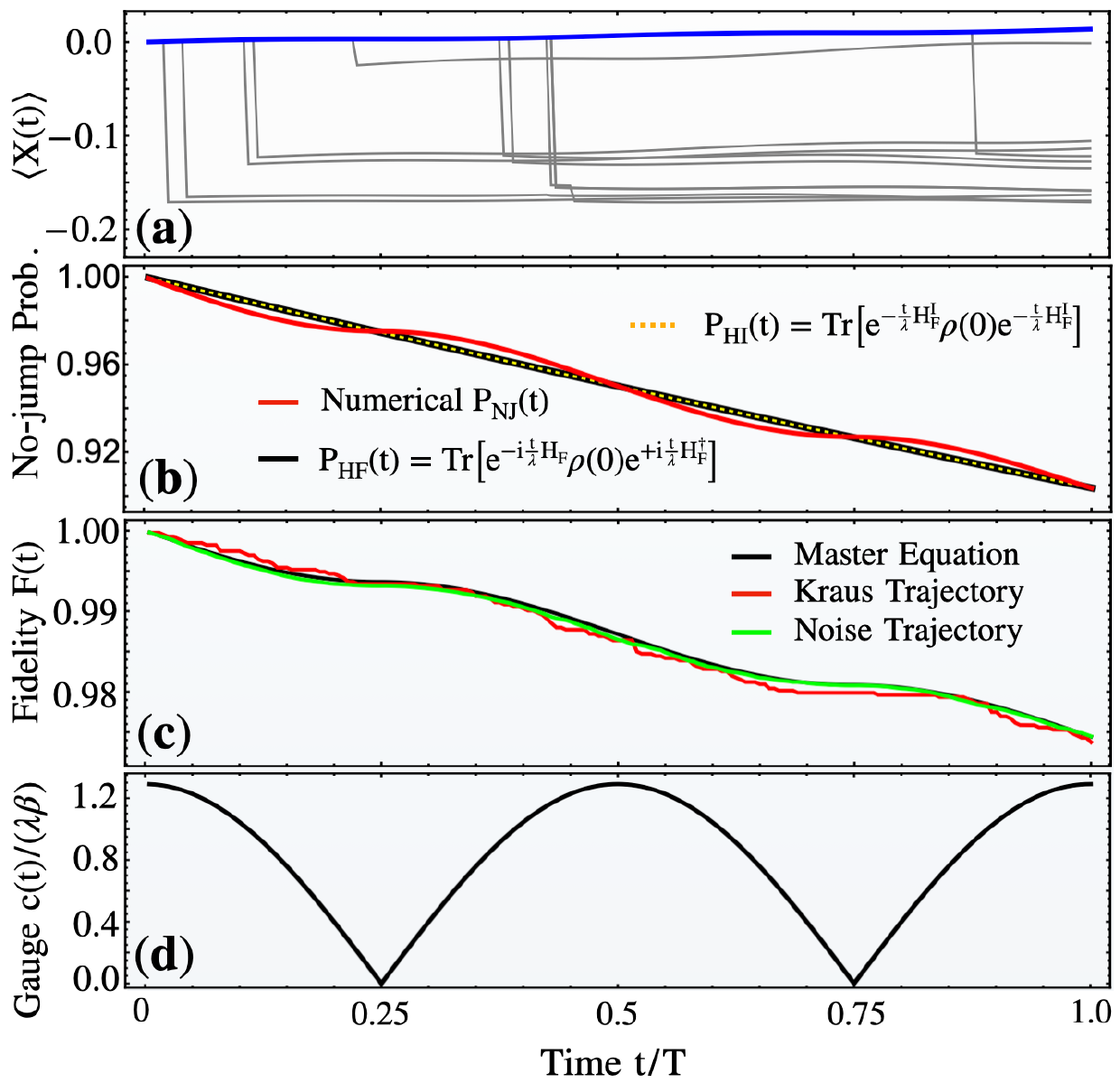}}
  \caption{{\bf Synthesizing NH cavity Hamiltonian via SFE.} {\bf (a)}  Monitoring quantum jumps of  Kraus trajectories generated from Eq.~(\ref{eq-rhozeta}) with the cavity quadrature $\langle X(t)\rangle$, where the trajectories with no quantum jumps are marked in blue. {\bf (b)}  Time-dependent no-jump probability given by numerical calculation (red), the Floquet Hamiltonian $H_F$ (blue curve), and the imaginary part of the Floquet Hamiltonian (yellow curve) with $H^I_F=\frac{1}{2i}(H^\dagger_F-H_F)$. {\bf (c)} Time-evolved fidelity $F(t)$ of density matrix $\rho(t)$ with respect to the initial state $\rho(0)=|0\rangle\langle 0|$ given by ensemble averaging Kraus trajectories (red), noisy trajectories (green), cf. Eq.~(\ref{eq-xit}), and the quantum master equation (black), cf. Eq.~(\ref{eq-drhot}). {\bf (d)} Time-dependent gauge term $c(t)$ that appears in Eq.~(\ref{eq-HstV}).
   }\label{Fig-benchmark}
\end{figure}

  \begin{figure*}
  \centering
 \centerline{\includegraphics[width=1.0\linewidth]{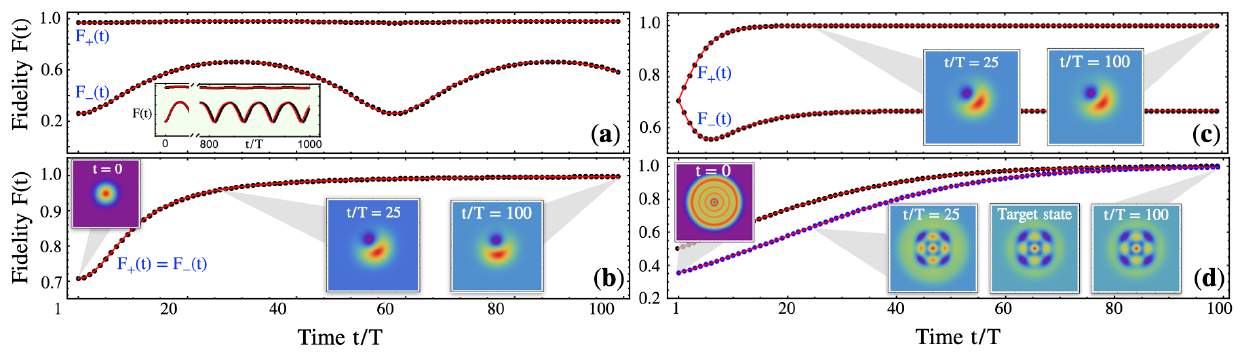}}
  \caption{{\bf Stroboscopic dynamcs of NH Hamiltonian.}{\bf (a)-(c):}
Stroboscopic time evolution of the fidelity $F_\pm(t)=\sqrt{{\rho^{1/2}(t)}\rho_\pm{\rho^{1/2}(t)}}$ with respect to the two eigenstates $\rho_\pm=|\psi_{\pm}\rangle|\psi_{\pm}|$ of the dissipative Hamiltonian (\ref{eq-disH}),  for the coupling parameter $\Gamma=0.5$ (a), $\Gamma=1.0$ (b), and $\Gamma=1.5$ (c). 
{\bf (d)} 
{ Time-evoluted fidelity $F(t)$ of the density matrix $\rho(t)$ generated by the designed NH Hamiltonia (\ref{eq-binomialHT}), with respect to the target kitten binomial state $|\psi_T\rangle=(|0\rangle+\sqrt{3}|4\rangle)/2$, from the initial groundstate $\rho(0)=|0\rangle\langle 0|$ (black dots) and the iinitial mixed state of $\rho(0)=n_c^{-1}\sum_{n=0}^{n_c-1}|n\rangle\langle n|$ with the truncated Fock nuumber $n_c=7$ (blue dots).}
We compare the stroboscopic dynamics by our SFE (black dots) to the results given by the target Hamiltonian  $\rho(t)=e^{-itH_T}\rho(0)e^{+itH^\dagger_T}$ (red solid curves), and show the Wigner functions of the density matrices at the selected time points as indicated in the plots. 
   }\label{Fig-stroboscopic}
\end{figure*}


\textit{Example.--}To verify our SFE scheme, we take the dissipative coupling Hamiltonian of a cavity in the Fock basis as an example
\bea\label{eq-disH}
H_T=\beta(|0\rangle\langle 0|-|1\rangle\langle 1|-i\Gamma |0\rangle\langle 1|-i\Gamma |1\rangle\langle 0|),\ \Gamma\in\mathbb{R}.\ \ 
\eea
The two nontrivial eignevalues of $H_T$ are $E_{\pm}=\pm\beta\sqrt{1-\Gamma^2}$ with two right eigenstates $|\psi_{\pm}\rangle=i\Gamma|1\rangle+(1\mp\sqrt{1-\Gamma^2})|0\rangle$ satisfying $H_T|\psi_{\pm}\rangle=E_{\pm} |\psi_{\pm}\rangle$.
The two eigenvalues are opposite real numbers for the coupling parameter $|\Gamma|<1$, but become two conjugate complex numbers for  $|\Gamma|>1$. At the exceptional point (EP) of $|\Gamma|=1$, $H_T$ has the twofold degeneracy of eigenvalues $E_{\pm}=0$, and the eigenstates coalesce into a single one.

 Given the target Hamiltonian (\ref{eq-disH}), the stochastic driving Hamiltonian (\ref{eq-HstV}) can be obtained from Eqs.~(\ref{eq-HRHI}), (\ref{eq-Vxt-cos}) and (\ref{eq-fT}).
We first benchmark the calculated stochastic periodic driving Hamiltonian that generates the Floquet Hamiltonian $(\ref{eq-HFc})$ over one Floquet period. To suppress the quantum jumps, we choose to monitor one quadrature of the cavity in the rotating frame, e.g., $X(t)={O}(t)x(t){O}^\dagger(t)$. In Fig.~\ref{Fig-benchmark}(a), we show the time evolution of the monitoring observable $\langle X(t)\rangle$ for one hundred example Kraus trajectories, where the abrupt changes indicate the occurrence of quantum jumps. In the weak driving regime ($\beta\ll 1$) that is needed for RWA, most trajectories do not encounter quantum jumps (blue curve). In Fig.~\ref{Fig-benchmark}(b), we track the time-dependent no-jump probability given by $P_{NJ}(t)=\prod_{n=0}^{[t/\Delta t]}\mathrm{Tr}[K_0(n\Delta t)\rho(n\Delta t) K^\dagger_0(n\Delta t)]$, where $\Delta t=T/N_t$ is the discrtized time step (red curve). For comparison, we also plot the no-jump probabilities given by the Floquet Hamiltonian $P_{FH}(t)=\mathrm{Tr}\big[e^{-i\frac{t}{\lambda}H_F}\rho(0)e^{+i\frac{t}{\lambda}H^\dagger_F}\big]$ (black curve) and $P_{FI}(t)=\mathrm{Tr}\big[e^{-\frac{t}{\lambda}H^I_F}\rho(0)e^{-\frac{t}{\lambda}H^I_F}\big]$ with $H^I_F=\frac{1}{2i}(H^\dagger_F-H_F)$ (orange dashed curve). The overlap of $P_{FH}(t)$ and $P_{FI}(t)$ indicates that the quantum jump happens mainly due to the imaginary part of the Floquet Hamiltonian. As the Floquet Hamiltonian $H_F$ describes the time-averaged dynamics over the Floquet period, there is some discrepancy between $P_{NJ}(t)$ and $P_{FH}(t)$ during the intermediate time ($0<t<T$). The calculated no-jump probabilities at the end of the Floquet period are consistent with the probability obtained by directly counting the Kraus trajectories, i.e., $P_{NJ}(T)\approx 0.97$ for $\beta=0.01$. 

We further calculate the density matrix $\rho(t)$ by ensemble averaging Kraus trajectories. In Fig.~\ref{Fig-benchmark}(c), we plot the time-evoluted fidelity of $\rho(t)$ with respect to the initial state $\rho(0)=|0\rangle\langle 0|$ (red), i.e., $F[\rho(t),\rho(0)]=\sqrt{\rho^{1/2}(t)\rho(0)\rho^{1/2}(t)}$, which is consistent with the results calculated by ensemble averaging noisy trajectories (green), cf. Eq.~(\ref{eq-xit}), and from the quantum master equation method (black), cf. Eq.~(\ref{eq-drhot}).
In Fig.~\ref{Fig-benchmark}(d), we plot the time-dependent gauge term $c(t)$ that appears in Eq.~(\ref{eq-HstV}), which is obtained by first calculating the eigenvalues of the operator $V_-(x,t)$ in the Fock basis, and then taking the opposite values of the minimum eigenvalues, ensuring the lowest eigenvalue of the operator $V_-(x,t)+c(t)$ keeps zero.

We then compare the stroboscopic dynamics using our SFE over multiple Floquet periods to the results predicted by the target NH Hamiltonian.
In Figs.~\ref{Fig-stroboscopic}(a)-(c), we calculate the stroboscopic time evolution of the fidelity with respect to the two eigenstates of the target Hamiltonian, i.e., $F_\pm(t)=\sqrt{{\rho^{1/2}(t)}\rho_\pm{\rho^{1/2}(t)}}$
with $\rho_\pm=|\psi_{\pm}\rangle|\psi_{\pm}|$,  for the coupling parameter $\Gamma=0.5$, $\Gamma=1.0$ and $\Gamma=1.0$, respectively. We also show the continuous time evolution of fidelities for the density matrix generated by the target Hamiltonian $\rho(t)=e^{-itH_T}\rho(0)e^{+itH^\dagger_T}$ (red solid curves). 
In Fig.~\ref{Fig-stroboscopic}(a), the fidelities exhibit oscillating behavior with stroboscopic time steps because the eigenvalues $E_\pm$ are both real numbers for $|\Gamma|<1$. In general, the fidelity time evolution generated by the stroboscopic dynamics of SFE agrees well with that generated by the NH target Hamiltonian. However, as shown by the inset, the discrepancy grows gradually due to the accumulation of non-RWA errors in the long time limit.
Fig.~\ref{Fig-stroboscopic}(b) shows that the two eigenstates coalesced into a single eigenstate with $E_\pm=0$ for the couping parameter $\Gamma=1.0$. In this case, the fidelity shows no oscillation but approaches the unit value in the long-time limit.  
Fig.~\ref{Fig-stroboscopic}(c) shows that the fidelity $F_-(t)$ approaches unit value in the long-time limit but the fidelity $F_+(t)$ approaches to the finite value of $\sqrt{\mathrm{Tr}[\rho_+\rho_-]}$ as the two eigenstates are non-orthogonal for the couping parameter $\Gamma=1.5$. We plot the Wigner functions of the density matrices at the selected time points as indicated in the plots. The asymptotic behaviour of the fidelity at the critical point $\Gamma=1$ shown in Fig.~\ref{Fig-stroboscopic}(b) follows a polynomial law, which is much slower than the exponential asymptotic behaviour for $|\Gamma|>1$ shown in Fig.~\ref{Fig-stroboscopic}(c).

\section{State purifying} 

The SFE provides a robust non-unitary scheme of preparing quantum states.
%
%
Given a target state $|\psi_T\rangle$, we first construct a set of states $\{|\psi_T^{(n)}\rangle | n \in \mathbb{N}, \langle \psi_T|\psi_T^{(n)}\rangle=0 \}$ such that all the states $|\psi_T^{(n)}\rangle$ ($n \in \mathbb{N}$) together with $|\psi_T\rangle$ form a complete orthogonal basis for the quantum system. Then, we set the target NH Hamiltonian as 
\bea\label{eq-binomialHT}
H_T=-i\gamma\sum_{n\in \mathbb{N}} |\psi_T^{(n)}\rangle\langle \psi_T^{(n)} |\ \ \ \mathrm{with}\ \ \ \gamma >0.
\eea 
Note that there is no target state component $|\psi_T\rangle\langle \psi_T|$ in the designed NH Hamiltonian. For example, to prepare one kitten binomial state $|\psi_T\rangle=(|0\rangle+\sqrt{3}|4\rangle)/2$, we can choose $\{|\psi_T^{(n)}\rangle | n \in \mathbb{N}\}=\{ |1\rangle,|2\rangle,  |3\rangle, (\sqrt{3}|0\rangle-|4\rangle)/2, |5\rangle,|6\rangle, \cdots\}$. 
In fact, all the state components orthogonal to the target state continuously decay, with only the target state remaining in the end. As a consequence, the cavity can be prepared to the target state from an \textit{arbitrary initial state} that has a finite overlap with the target state. {In Fig.~\ref{Fig-stroboscopic}(d), we calculate the time-evoluted fidelity with respect to the target kitten binomial state from two different initial states, i.e, the ground state $\rho(0)=|0\rangle\langle 0|$ and the mixed state of $\rho(0)=\frac{1}{n^c}\sum_{n=0}^{n_c-1}|n\rangle\langle n|$ with $n_c$ the truncation Fock number.
We plot the Wigner functions of the density matrix at the selected time points, along with those of the target state. The choice of an arbitrary initial state indicated that the SFE state preparation method is robust to quantum errors.

\section{Disscussions}
%
One key ingredient of implementing the SFE is to monitor the quantum jumps by designing a proper measurement scheme. The basic idea is to make the state of the measurement setup nearly frozen for no-jump trajectories, but dramatically altered when quantum jumps occur.
For the example of superconducting cavity, we can choose the qudrature current $I_s$ as the monitoring observable, and couple it to the Josephson bifurcation amplifier (JBA) resonator~\cite{siddiqi2004prl,siddiqi2006prb} with Hamiltonian $H_{\mathrm{JBA}}=Q^2/2C-E_J\cos\delta-\hbar (2e)^{-1}( I_0+I_s)\delta$, where $Q$, $\delta$, $C$ and $E_J$ are the charge, phase difference, capacitor and Josephson energy of the JBA resonator. The bias current $I_0$ is used to control the bistability of JBA with small amplitude state (SAS) and large amplitude state (LAS). We can set the parameters such that  JBA stays in the SAS when there is no quantum jump, while it becomes LAS when quantum jumps occur.
Note that it does not need to identify when exactly the quantum jumps occurred, but only to measure the accumulated effects of quantum jumps at the end of one Floquet period. 

The current work does not consider the loss or gain of quantum systems, nor other practical facts in the experiments. Instead, we aim to provide a theoretical framework, with no need for loss or gain, to synthesize arbitrary NH Hamiltonians for general quantum systems by extending traditional Floquet engineering to the \textit{stochastic Floquet engineering} with merely Hermitian drives. In principle, arbitrary non-unitary dynamics can be generated via quantum trajectories without quantum jumps. Our SFE method takes noise as a valuable quantum resource that can be engineered for NH physics and non-unitary quantum computing. Future works will investigate the interplay between noise and dissipation that has drawn much attention recently~\cite{wang2026arxiv,yang2026arxiv}.
 }


\bigskip

\begin{acknowledgements}
This work is supported by the NSFC (Grant No. 12421005),  the National Natural Science Foundation of China (Grant No. 12475025), the Hunan Major Sci-Tech Program (2023ZJ1010), and the Innovation Program for Quantum Science and Technology (2024ZD0301000).
\end{acknowledgements}

\newpage

\onecolumngrid
\appendix

\section{General Theory}\label{}

\subsection{Stochastic Hamiltonian}

Suppose the dynamics of a quantum system is given by 
$
{d\rho}/{dt}=-i\lambda^{-1}[H(t),\rho]\equiv\mathcal{D}(t)\rho,
$
where $\rho$ is the density operator of the system, $H(t)$ is the Hamiltonian and $\lambda$ is the dimensionless Planck constant.  The time evolution of the density matrix is 
$
\rho(t)=U(t)\rho(t_0) U^\dagger(t),
$
where $U(t)$ is the time evolution operator given by
\bea
U(t,t_0)=\mathcal{T}\exp\big[-\frac{i}{\lambda}\int_{t_0}^{t}{H}(\tau)d\tau\big]
\eea
with $\mathcal{T}$ the time-ordering operator. 
Now we add a stochastic Hamiltonian term $\sqrt{\eta}\xi(t)H'(t)$, where $\xi(t)$ is the white noise that follows $\langle{\xi(t)}\rangle=0$ and $\langle{\xi(t)\xi(t')}\rangle=\delta(t-t')$. Then, the dynamics of the quantum system with the total Hamiltonian are  
\bea
\frac{d\rho_{\xi}}{dt}=-\frac{i}{\lambda}[H(t)+\sqrt{\eta}\xi(t)H'(t),\rho_{\xi}]\equiv\mathcal{D}(t)\rho_{\xi}+\sqrt{\eta}\xi(t)\mathcal{K}(t)\rho_{\xi},
\eea
where we have defined another superoperator $\mathcal{K}(t)\rho_{\xi}\equiv-i\lambda^{-1}[H'(t),\rho_{\xi}]$ to distinguish the superoperator $\mathcal{D}(t)$. Now, the time evolution of the density matrix conditioned on the noise process is
$
\rho_\xi(t)=U_\xi(t)\rho(t_0) U_\xi^\dagger(t)
$
with the conditioned time evolution operator given by
\bea
U_\xi(t,t_0)=\mathcal{T}\exp\Big(-i\frac{1}{\lambda}\int_{t_0}^{t}\big[{H}(\tau)+\sqrt{\eta}\xi(\tau)H'(\tau)\big]d\tau\Big).
\eea
In an infinitesimal time interval $\Delta t$, the density operator becomes
\bea
\rho_{\xi}(t+\Delta t)&=&U_\xi(t+\Delta t,t)\rho(t) U_\xi^\dagger(t,t+t_0)\nl
&=&e^{(\mathcal{D}+\sqrt{\eta}\xi\mathcal{K})\Delta t}\rho_{\xi}(t) \nl
&\approx&e^{\sqrt{\eta}\xi{\mathcal{K}}\Delta t}e^{\mathcal{D}\Delta t}\rho_{\xi}(t) \nl
&\approx&(1+\sqrt{\eta}\xi{\mathcal{K}}\Delta t+\frac{1}{2}\eta\xi^2{\mathcal{K}}^2\Delta t^2)(1+\Delta t\mathcal{D})\rho_{\xi}(t)\nl
&=&(1+\sqrt{\eta}\Delta W\mathcal{K}+\frac{1}{2}\eta\Delta W^2\mathcal{K}^2)(1+\Delta t\mathcal{D})\rho_{\xi}(t)\nl
&\approx&\Big[1+\Delta t\mathcal{D}(t)+\frac{1}{2}\eta\Delta W^2\mathcal{K}^2(t)+\sqrt{\eta}\Delta W\mathcal{K}(t)\Big]\rho_{\xi}(t).
\eea
 Here, we have introduced the Wiener increasement $\Delta W\equiv \xi(t)\Delta t$, which follows $\langle {\Delta W}\rangle=0$ and  $\langle {\Delta W^2}\rangle=\Delta t$. By keeping the terms up to the order of $\Delta t$, we have
\bea
\rho_{\xi}(t+\Delta t)&=&\rho_{\xi}(t)+\Delta t\mathcal{D}\rho_{\xi}+\frac{1}{2}\eta\Delta W^2\mathcal{K}^2\rho_{\xi}+\sqrt{\eta}\Delta W\mathcal{K}\rho_{\xi}
\eea
We require that the white noise process $\xi(t)$ is much faster than the engineered Hamiltonians $H(t)$ and $H'(t)$.
Assuming that $N_{\xi}$ discretized random numbers are evenly generated during the time step $\Delta t$, we have the time evolution of the density matrix with the infinitesimal time interval $\Delta t/N_\xi$ as follows
\bea
\rho_{\xi}\Big(t+\frac{\Delta t}{N_\xi}\Big)&=&\Big[1+\frac{\Delta t}{N_\xi}\mathcal{D}+\frac{1}{2}\eta\Delta W_0^2\mathcal{K}^2+\sqrt{\eta}\Delta W_0\mathcal{K}\Big]\rho_{\xi}(t)\nl
\rho_{\xi}\Big(t+2\frac{\Delta t}{N_\xi}\Big)&=&\Big[1+\frac{\Delta t}{N_\xi}\mathcal{D}+\frac{1}{2}\eta\Delta W_1^2\mathcal{K}^2+\sqrt{\eta}\Delta W_1\mathcal{K}\Big]\rho_{\xi}\Big(t+\frac{\Delta t}{N_\xi}\Big)\nl
&\vdots&\nl
\rho_{\xi}\Big(t+\Delta t\Big)&=&\Big[1+\frac{\Delta t}{N_\xi}\mathcal{D}+\frac{1}{2}\eta\Delta W_{N_\xi-1}^2\mathcal{K}^2+\sqrt{\eta}\Delta W_{N_\xi-1}\mathcal{K}\Big]\rho_{\xi}\Big(t+(N_\xi-1)\frac{\Delta t}{N_\xi}\Big).
\eea
Here, we have divided the Wiener increasement into $\Delta W=\int_{t_0}^{t_0+\Delta t}\xi(\tau)d\tau\approx \sum_{k=1}^{N_{\xi}}\Delta W_k$, where $\Delta W_k\equiv \xi(t_0+k\Delta t)\Delta t/N_{\xi}$ are Gaussian random numbers with zero mean value and variance $\Delta t/N_{\xi}$. 
We summarize the above equations and get the state of the density matrix after a time interval $\Delta t$ is just the average of many intermediate density matrices, i.e., 
\bea\label{eq-rhoxit}
\rho_{\xi}(t+\Delta t)-\rho_\xi(t)&=&\sum_{k=0}^{N_\xi-1} \Big[\frac{\Delta t}{N_\xi}\mathcal{D}+\frac{1}{2}\eta\Delta W_k^2\mathcal{K}^2+\sqrt{\eta}\Delta W_k\mathcal{K}\Big]\rho_{\xi}\Big(t+k\frac{\Delta t}{N_\xi}\Big).
\eea
By replacing the stochastic density matrix $\rho_\xi(t)$ by the averaged value 
$\rho(t)\approx \sum_{k=0}^{N_\xi-1}\rho_{\xi}\Big(t+k\frac{\Delta t}{N_\xi}\Big)=\langle \rho_\xi(t)\rangle$ and taking the limit $\Delta t\rightarrow 0$, 
we have dynamics for the density matrix $\rho(t)$ from Eq.~(\ref{eq-rhoxit}) as follows
\bea\label{eq-drho}
\frac{d\rho}{dt}&=&\mathcal{D}\rho+\frac{1}{2}\eta\mathcal{K}^2\rho\nl
&=&\mathcal{D}\rho+\frac{1}{2\lambda^2}\eta[-iH',-i[H',\rho]]\nl
&=&-i\frac{1}{\lambda}[H(t),\rho]+\eta\frac{1}{\lambda^2} \big(H'\rho H'-\frac{1}{2}H'^2\rho-\frac{1}{2}\rho H'^2\big)\nl
&=&-i \frac{1}{\lambda}\big[\mathcal{H}(t)\rho-\rho \mathcal{H}^\dagger(t)\big]+ \frac{\eta}{\lambda^2}H'\rho H'.
\eea
In summary, the original stochastic Hamiltonian $\sqrt{\eta}\xi(t)H'$ produces an imaginary Hamiltonian $-i\eta H'^2/2$ together with a quantum jump term $\eta H'\rho H'$.
Neglecting the quantum jump terms, the system is described by the following non-Hermitian Hamiltonian 
\bea
\mathcal{H}(t)=H(t)-i\frac{1}{2}\frac{\eta}{\lambda} H'^2(t).
\eea

\subsection{Floquet dynamics}

A periodically driven system is also called a \textit{Floquet system}~\cite{Floquet1883,Shirley1965pr}. For the time-periodc Hamiltonians $H(t+T)=H(t)$ and $H'(t+T)=H'(t)$, the density matrix during one time period is given by
\bea
\rho_{\xi}(t_0+T)&\approx&\prod_{n=1}^{T/\Delta t} e^{[\mathcal{D}(t_0+n\Delta t)+\sqrt{\eta}\xi{\mathcal{K}}(t_0+n\Delta t)]\Delta t}\rho_{\xi}(t_0) \nl
&\approx&\prod_{n=1}^{T/\Delta t}\Big[1+\Delta t\mathcal{D}(t_0+n\Delta t)+\frac{1}{2}\eta\Delta W^2_n\mathcal{K}^2(t_0+n\Delta t)+\sqrt{\eta}\Delta W_n\mathcal{K}(t_0+n\Delta t)\Big]\rho_{\xi}(t_0) \nl
&\approx&\rho_{\xi}(t_0)+\sum_{n=1}^{T/\Delta t}\Big[\Delta t\mathcal{D}(t_0+n\Delta t)+\frac{1}{2}\eta\Delta W^2_n\mathcal{K}^2(t_0+n\Delta t)+\sqrt{\eta}\Delta W_n\mathcal{K}(t_0+n\Delta t)\Big]\rho_{\xi}(t_0)\nl
&&+\sum_{n\neq n'}^{T/\Delta t}\eta\Delta W_n\Delta W_{n'}\mathcal{K}(t_0+n\Delta t)\mathcal{K}(t_0+n'\Delta t)\rho_{\xi}(t_0)+\cdots.
\eea
Here, we have neglected the expansion terms higher than the order of $\Delta t$.
Using $\langle {\Delta W_n}\rangle=0$, $\langle {\Delta W^2_n}\rangle=\Delta t$, and  $\langle {\Delta W_n \Delta W_{n'}}\rangle=\langle {\Delta W_n\rangle\langle \Delta W_{n'}}\rangle=0$ for $n\neq n'$, we have time evolution for the averaged density matrix $\rho(t)=\langle \rho_\xi(t)\rangle$, 
\bea
\rho(t_0+T)&=&\rho(t_0)+\sum_{n=1}^{T/\Delta t}\Big[\Delta t\mathcal{D}(t_0+n\Delta t)+\frac{1}{2}\eta\Delta t\mathcal{K}^2(t_0+n\Delta t)\Big]\rho(t_0)\nl
&=&\rho(t_0)+\Big[\int_{t_0}^{t_0+T} \mathcal{D}(t)dt+\frac{1}{2}\eta \int_{t_0}^{t_0+T} \mathcal{K}^2(t)dt\Big]\rho(t_0)\nl
&=&\rho(t_0)+T \overline{\mathcal{D}}\rho(t_0)+\frac{T}{2}\eta \overline{\mathcal{K}^2}\rho(t_0).
\eea
where we have defined the time average for a periodic operator $\overline{O}=T^{-1}\int_{t_0}^{t_0+T} {O}(t)dt$. From Eq.~(\ref{eq-drho}),  we have the master equation for the strobascopic dynamics
\bea\label{eq-1}
\frac{\Delta\rho}{\Delta t}=\frac{\rho(t_0+T)-\rho(t_0)}{T}
&=&\overline{\mathcal{D}}\rho-i \frac{1}{\lambda^2}\big[(-i\frac{1}{2}\eta \overline{H'^2})\rho-\rho(-i\frac{1}{2}\eta \overline{H'^2})\big]+\eta\frac{1}{\lambda^2} \overline{H'\rho H'}\nl
&=&-i\frac{1}{\lambda} \big[H_F\rho(t)-\rho(t) H^\dagger_F\big]+\eta\frac{1}{\lambda^2} \overline{H'\rho H'}\nl
&\equiv&-i\frac{1}{\lambda}\mathcal{L}(t_0)\rho.
\eea
where $H_F$ is the effective Floquet Hamiltonian given by
\bea\label{eq-HFrho}
H_F\equiv\overline{H(t)}-i\frac{\eta}{2\lambda} \overline{H'^2},
\eea
and $\mathcal{L}(t_0)$ is the defined unconditional superoperator. 
For single-trajectory dynamics, the time evolution of the conditional density matrix is given by
$
\rho_\xi(t)=U_\xi(t,t_0)\rho(t_0) U_\xi^\dagger(t,t_0)
$
from the unconditional density matrix $\rho(t_0)$.
Therefore, we can write strobascopic dynamics in the Fock representation as
\bea
\frac{\langle n|\rho_\xi(T)-\rho(t_0)|m\rangle}{T}&=&\langle n|U_\xi(t_0+T,t_0)\rho(t_0) U_\xi^\dagger(t_0+T,t_0)-\rho(t_0)|m\rangle\nl
&=&\sum_{n',m'}\Big(\langle n|U_\xi(t_0+T,t_0)|n'\rangle \langle m'| U_\xi^\dagger(t_0+T,t_0)|m\rangle-\delta_{nn'}\delta_{mm'}\Big) \langle n'|\rho(t_0)|m' \rangle\nl
&\equiv&-i\frac{1}{\lambda}\sum_{n',m'}\mathcal{L}^{nm,n'm'}_\xi(t_0) \langle n'|\rho(t_0)|m' \rangle.
\eea
Here, we have introduced the conditional (stochastic) superoperator $\mathcal{L}_\xi(t_0)$ whose matrix element is given by
\bea
\mathcal{L}^{nm,n'm'}_\xi(t_0) &\equiv&i\lambda\langle n|U_\xi(t_0+T,t_0)|n'\rangle \langle m'| U_\xi^\dagger(t_0+T,t_0)|m\rangle-i\lambda\delta_{nn'}\delta_{mm'}.
\eea
The unconditional superoperator $\mathcal{L}(t_0)$ is the average of  $\mathcal{L}_\xi(t_0)$, i.e., $\mathcal{L}(t_0)=\langle\mathcal{L}_\xi(t_0)\rangle$.
According to Eq.~(\ref{eq-1}), the matrix element of the unconditional superoperator $\mathcal{L}(t_0)$ is given by
\bea
\mathcal{L}^{nm,n'm'}(t_0)
 &=&\langle n|H_F|n'\rangle \delta_{mm'}-\langle m'|H^\dagger_F|m\rangle \delta_{nn'}+ i\frac{\eta}{\lambda}\overline{\langle n|H'(t)|n'\rangle\langle m'|H'(t)|m\rangle}
\eea

\subsection{Procedure to engineering non-Hermitian Hamiltonian}
Here, we provide a brief summary of the procedure to engineering a target non-Hermitian Hamiltonian. First, given a target non-Hermitian Hamiltonian $H_T=H_R-iH_I$, we can identify its real part and imaginary part by
\bea\label{eq-2}
H_R=\frac{1}{2}(H_T^{\dagger}+H_T),\ \ \ \ H_I=\frac{1}{2i}(H^\dagger_T-H_T)
\eea

Second, we set $\eta=2\lambda$ and engineer two periodic Hermitian Hamiltonians $H(t)$ and $h(t)$ such that 
\bea\label{eq-3}
\overline{H(t)}\equiv\frac{1}{T}\int_0^TH(t)dt=\lambda H_R, \ \ \ \overline{h(t)}\equiv\frac{1}{T}\int_0^Th(t)dt=\lambda H_I.
\eea

Third, we engineer a Hermitian Hamiltonian with a stochastic part 
\bea
\mathcal{H}(t)=H(t) + \sqrt{2\lambda}\xi(t) H'(t),\ \ \ \ \mathrm{with}\ \ \ H'(t)=\sqrt{h(t)+c(t)I} ,
\eea
where the identity matrix $I$ with $c(t)$ is a free gauge to guarantee the psotivity of $h(t)+c(t)I$. According to Eqs.~(\ref{eq-1}), (\ref{eq-2}) and (\ref{eq-3}), we have the effective Hamiltonian of Floquet dynamics
\bea\label{eq-HFic}
H_F=\lambda (H_R-iH_I)-i\overline{c(t)}=\lambda H_T-i\overline{c}.
\eea
The master equation (\ref{eq-1}) is given by 
\bea\label{}
\frac{\Delta\rho}{\Delta t}=
&=&-i\frac{1}{\lambda} \big[H_F\rho(t)-\rho(t)H^\dagger_F\big]+\frac{2}{\lambda} \overline{\sqrt{h(t)+c(t)I}\rho \sqrt{h(t)+c(t)I}}.
\eea
Note that the imaginary constant in the effective Hamiltonian (\ref{eq-HFic}) cannot be simply neglected because its commutator in the master equation is not zero but $[-i\overline{c},\rho]=-2i\overline{c}\rho$, resulting in a decay term $-2\overline{c}\rho$.

\section{Applications of Cavity NH Hamiltonian Engineering}

\subsection{Arbitrary Phase-space Hamiltonian engineering}

In order to generate the target Hamiltonian $\hat{H}_T$, which is in general an arbitrary function of quadrature operators $\hat{x}$ and $\hat{p}$, we drive the cavity by a periodic external potential $V(\hat{x},t)=V(\hat{x},t+T_d)$ with $T_d=2\pi/\omega_d$, i.e.,
\begin{equation}
\hat{\mathcal{H}}(t)=\frac{\omega_0}{2}\left(\hat{p}^{2}+\hat{x}^{2}\right)+\beta V(\hat{x},t).
\label{eq-machcalHt}
\end{equation}
A periodically driven system is also called a \textit{Floquet system}~\cite{Floquet1883,Shirley1965pr}. 
By transforming the above Hamiltonian into the rotating frame of frequency $\Omega=2\pi/T$ with $T=nT_d $ $(n\in \mathbb{Z}^+)$, we have 
$\hat{O}(t)\hat{x}\hat{O}^\dagger(t)=\hat{x}\cos (\Omega t)+\hat{p}\sin (\Omega t)$  with time-evolution operator $\hat{O}(t)\equiv e^{i\hat{a}^\dagger\hat{a}\Omega t}$. The transformed Hamiltonian in the rotating frame is given by
\bea\label{eq-Ht}
\hat{H}(t)&\equiv&\hat{O}(t)\hat{\mathcal{H}}(t)\hat{O}^\dagger(t)-i\lambda \hat{O}(t)\dot{\hat{O}}^\dagger(t)\nl
&=& \beta V\Big[\hat{x}\cos (\Omega t)+\hat{p}\sin (\Omega t),t\Big].
\eea
Here,  we have adapted the multi-photon resonance condition $T=2\pi/\omega_0$ or equivalently $\Omega=\omega_0$, i.e., the driving frequency is set to be $n$ times the bare frequency of the harmonic oscillator.

The Flouqet theorem states that the stroboscopic time evolution of a periodic time-varying system is described by a time-independent Floquet Hamiltonian $\hat{H}_F$ determined by  \cite{Floquet1883,Shirley1965pr,Sambe1973pra,Grifoni1998pr,Eckardt2015NJP,Liang2018njp}
\bea\label{eq-HFt0}
\exp\Big(-i\frac{1}{\lambda}\hat{H}_FT\Big)=\mathcal{T}\exp\Big[-i\frac{1}{\lambda}\int_{0}^{T}\hat{H}(t)dt\Big], 
\eea
where $\mathcal{T}$ is the time-ordering operator. 
Under the rotating wave approximation (RWA), the Floquet Hamiltonian $\hat{H}_F$ is just the time-averaged version of $\hat{H}(t)$ over one Floquet period $T$ ~\cite{guo2024prl,Eckardt2015NJP,Mikami2016prb}, i.e.,
\bea\label{eq-h0h1h3}
\lim_{\omega_0/\beta\rightarrow\infty}\hat{H}^{}_F(\hat{x},\hat{p})&=&\frac{1}{T}\int_{0}^{T}dt  \hat{H}(t).\ \ \ 
\eea
By properly engineering the driving potential $V(\hat{x},t)$~\cite{guo2024prl}, the Floquet Hamiltonian $\hat{H}_F(\hat{x},\hat{p})$ can be designed as the target Hamiltonian $\hat{H}_T(\hat{x},\hat{p})$.

For this purpose, we decompose a given target Hamiltonian $\hat{H}_T(\hat{x},\hat{p})$ as a sum of plane-wave operators in the noncommutative phase space~\cite{guo2024prl}, i.e.,
\bea\label{eq-HTxp}
\hat{H}_T(\hat{x},\hat{p})
=
\frac{1}{2\pi}\int \int dk_x dk_pf_T(k_x,k_p)e^{i(k_x\hat{x}+k_p\hat{p})},
\eea
where the \textit{noncommutative Fourier transformation} (NcFT) coefficient in Eq.~(\ref{eq-HTxp}) is given by~\cite{guo2024prl}
\bea\label{eq-fFT}
f_T(k_x,k_p)=\frac{e^{\frac{\lambda}{4}(k^2_x+k^2_p)}}{2\pi}\int\int dxdp H^Q_T(x,p) e^{-i(k_xx+k_pp)}.\nl 
\eea
Here, the integrand $H^Q_T(x,p)=\langle \alpha |\hat{H}_T|\alpha \rangle$ is the Q-function of the target Hamiltonian with $|\alpha\rangle$ the coherent state defined via $\hat{a}|\alpha\rangle=\alpha |\alpha\rangle$, where $\alpha=(x+ip)/\sqrt{2\lambda}$ with $x\equiv\langle \alpha |\hat{x}|\alpha \rangle$ and $p\equiv\langle \alpha |\hat{p}|\alpha \rangle$.

With the NcFT coefficient, one can design the driving potential by superposing a series of cosine-type lattice potentials as~\cite{guo2024prl}
\bea\label{eq-Vxt-2}
V(x,\Omega t)
&=&\int_{-\infty}^{+\infty}A(k, \Omega t)\cos[kx+\phi(k,\Omega t)]dk.\ \ 
\eea
Here, the tunable time-dependent amplitude $A(k,\Omega t)$ and phase $\phi(k,\Omega t)$ are given by
\bea\label{eq-Aphi}
 \left \{ \begin{array}{lll}
 A(k,t)&=&k\Big|f_{T}(k\cos{\Omega t},k\sin{\Omega t})\Big|\\
\phi(k,t)&=&\text{Arg}\Big[f_{T}(k\cos{\Omega t},k\sin{\Omega t})\Big],
\end{array} \right.
\eea
where we have adopted $k_{x}=k\cos{\Omega t}$ and $k_{p}=k\sin{\Omega t}$. 
Each cosine component can be implemented with, e.g., an optical lattice that is formed by laser beams intersecting at an angle in cold-atom experiments \cite{Moritz2003prl,Hadzibabic2004prl,Guo2022prb} or a JJ potential in superconducting circuits \cite{Chen2014prb,Hofheinz2011prl,Chen2011apl}.
Note that, according to the definitions given in  Eqs.~(\ref{eq-machcalHt}), (\ref{eq-Ht}), (\ref{eq-h0h1h3}), (\ref{eq-HTxp}), and (\ref{eq-Vxt-2}),  the Floquet and target Hamiltonians actually differ by an overall prefactor, i.e., $\hat{H}_F=\beta \hat{H}_T$.

\subsection{Example }
We aim to generate the target Hamiltonian of a cavity with the selected two Fock basis of $|n\rangle, |m\rangle$ given by
\bea
H_T=\sum\limits_{n',m'\in \{n,m\}}c_{n'm'}|n'\rangle\langle m'|,
\eea
where the Hamiltonian is non-Hermitian $c_{nm}\neq c^*_{mn}$. Then, we have the real part $H_R=\frac{1}{2}(H_T^{\dagger}+H_T)$ and the imaginary part $H_I=\frac{1}{2i}(H^\dagger_T-H_T)$ given by
\bea\label{}
 \left \{ \begin{array}{lll}
H_R&=&\sum\limits_{n',m'\in \{n,m\}}\frac{1}{2}\big(c^*_{m'n'}+c_{n'm'}\big)|n'\rangle\langle m'|\\
 H_I&=&\sum\limits_{n',m'\in \{n,m\}}\frac{1}{2i}\big(c^*_{m'n'}-c_{n'm'}\big)|n'\rangle\langle m'|.
\end{array} \right.
\eea
In order to generate the target Hamiltonian $\hat{H}_T$, we drive the cavity by two periodic driving potentials, i.e.,
\begin{equation}\label{eq-machHt}
\hat{\mathcal{H}}(t)=\frac{\omega_0}{2}\left(\hat{p}^{2}+\hat{x}^{2}\right)+\beta U (\hat{x},t)+ \sqrt{2\beta}\xi(t)\sqrt{h(x,t)+c} .
\end{equation}
We construct  the driving potential by superposing a series of cosine-type lattice potentials as~\cite{guo2024prl}
\bea\label{eq-Vxt-0}
V(x, t)
&=&\int_{-\infty}^{+\infty}A(k, t)\cos[kx+\phi(k, t)]dk.\ \ 
\eea
Here, the tunable time-dependent amplitude $A(k, t)$ and phase $\phi(k, t)$ are given by
 $A(k,t)=k\big|f_{T}(k,\omega_0 t)\big|$ and $\phi(k,t)=\text{Arg}\big[f_{T}(k,\omega_0 t)\big].$
%
The noncommutative Fourier coefficient of the target Hamiltonian is given by
\bea\label{eq-fTdef}
f_{T}(k,\omega_0 t)&=&\sum_{n',m'\in \{n,m\}}\tilde{c}_{n',m'}f_{n',m'}(k,\omega_0 t)
\eea
where each NcFT component  is given by $f_{n'm'}(k,\omega_0 t)=\sqrt{\frac{n'!}{m'!}}\left(\frac{i}{k}\sqrt{\frac{2}{\lambda}}\right)^{m'-n'}\frac{\lambda e^{\frac{\lambda}{4}k^{2}+i(m'-n')\omega_0t}}{\Gamma(1+n'-m')}{}_{1}F_{1}(1+n'; 1+n'-m'; -\frac{\lambda}{2}k^{2})$. Here, $J_{n'-m'}(z)$ is the Bessel function of the first kind (of order $n'-m'$),
$\Gamma(n)$ is the Gamma function, and ${}_{1}F_{1}(a;b;z)$ is the Kummer confluent hypergeometric function.

The driving potential $U(x,t)$ and the function $h(x,t)$ in Eq.~(\ref{eq-machHt}) are given by $V(x,t)$ of Eq.~(\ref{eq-Vxt-0}) by taking  $\tilde{c}_{n',m'}=(c^*_{m'n'}+c_{n'm'})/2$ and $\tilde{c}_{n',m'}=(c^*_{m'n'}-c_{n'm'})/2i$ in Eq.~(\ref{eq-fTdef}) respectively.
We transform the above Hamiltonian into the rotating frame with time-evolution operator $\hat{O}(t)\equiv e^{i\hat{a}^\dagger\hat{a}\omega_0 t}$, i.e., 
$
\tilde{\mathcal{H}}(t)\equiv\hat{O}(t)\hat{\mathcal{H}}(t)\hat{O}^\dagger(t)-i\lambda \hat{O}(t)\dot{\hat{O}}^\dagger(t).
$
According to the NcFT method, the effective Floquet Hamiltonian~(\ref{eq-HFrho}) that describes the stroboscopic dynamics of the driven cavity is given by
\bea\label{}
H_F=\frac{1}{T}\int_0^T\tilde{\mathcal{H}}(t)dt=\beta(H_R-iH_I-i\overline{c})=\beta(H_T-i\overline{c})
\eea
together with a quantum jump term in the master equation (\ref{eq-1})
given by
$
2\beta \overline{\sqrt{h(t)+\overline{c}I}\rho \sqrt{h(t)+\overline{c}I}}.
$

\end{document}